\documentclass[12pt]{article}
\pdfoutput=1
\usepackage{geometry,enumerate,amsmath,amssymb}
\usepackage{fullpage}
\usepackage{graphicx}
\numberwithin{equation}{section}

\newcommand{\be}{\begin{equation}}
\newcommand{\ee}{\end{equation}}
\newcommand{\bea}{\begin{eqnarray}}
\newcommand{\eea}{\end{eqnarray}}

\renewcommand{\hat}{\widehat}

\renewcommand{\epsilon}{\varepsilon}
\begin{document}
\title{
  Phases of kinky holographic nuclear matter
}
\author{Matthew Elliot-Ripley,
  Paul Sutcliffe and Marija Zamaklar\\[10pt]
  {\em \normalsize
    Department of Mathematical Sciences,}
{\em \normalsize  Durham University, Durham DH1 3LE, U.K.}\\[10pt]
{\normalsize   m.k.i.d.elliot-ripley@durham.ac.uk}\\
{\normalsize p.m.sutcliffe@durham.ac.uk}\\
{\normalsize marija.zamaklar@durham.ac.uk}\\[10pt]
}
\date{July 2016}
\maketitle
\begin{abstract}
  Holographic QCD at finite baryon number density and zero temperature is studied within the five-dimensional Sakai-Sugimoto model. We introduce a new approximation that models a smeared crystal of solitonic baryons by assuming spatial homogeneity to obtain an effective kink theory in the holographic direction.
  The kink theory correctly reproduces a first order phase transition to lightly bound nuclear matter. As the density is further increased the kink splits into a pair of half-kink constituents, providing a concrete realization of the previously
 suggested dyonic salt phase, where the bulk soliton splits into constituents at high density. The kink model also captures the phenomenon of
 baryonic popcorn, in which a first order phase transition generates
 an additional soliton layer in the holographic direction.
 We find that this popcorn transition takes place at a density below
the dyonic salt phase, making the latter energetically unfavourable.
 However, the kink
 model predicts only one pop, rather than the sequence of
 pops suggested by previous approximations.
 In the kink model the two layers   
 produced by the single pop form the surface of a soliton bag that increases in size as the baryon chemical potential is increased. The interior of the bag is filled with abelian electric potential and the instanton charge density is localized on the surface of the bag.
  The soliton bag may provide a holographic description of a quarkyonic phase.

  \end{abstract}

\newpage
\section{Introduction}\quad
Determining the properties and phases of cold and dense nuclear matter, as found for example in the interior of neutron stars, is a difficult task, because neither lattice methods nor perturbation theory can be applied in this regime of QCD.
Holographic methods allow the study of strongly-coupled QCD-like theories as a function of temperature and baryon chemical potential, to hopefully provide some useful qualitative information about the rich variety of phases and transitions that might be expected in QCD.
Pioneering studies \cite{Kim:2006gp,Bergman:2007wp,Rozali:2007rx}
on cold and dense nuclear matter in holographic QCD have used the  
Sakai-Sugimoto model
\cite{Sakai:2004cn,Sakai:2005yt}, as this is the preeminent example of a string theory realization of a holographic QCD-like theory.
The two flavour version has a low energy description as a five-dimensional $U(2)$ gauge theory with a Chern-Simons term, in which baryons correspond to topological solitons in the bulk.
As baryon number
is identified with the instanton number of the gauge field, the study of baryons
requires the construction of Yang-Mills-Chern-Simons solitons in curved space with a prescribed instanton number.

It is a significant challenge to calculate the required solitons, even numerically, and to date only the single baryon solution has been computed \cite{Bolognesi:2013nja}.
The computation in the baryon number one sector is facilitated by the
application of continuous symmetries that are not present for higher baryon numbers, or at finite baryon number density. As a consequence, there
are currently no numerical solutions available at finite density, either in the full Sakai-Sugimoto model or its five-dimensional version.
This absence of solutions has motivated the use of a number of
approximate methods
\cite{Rozali:2007rx,Kim:2007vd,Kaplunovsky:2012gb,Li:2015uea}
to study the Sakai-Sugimoto model at finite
density. Often this involves the use of point-like instantons, but below we
discuss some alternative approaches that share some common features with the effective kink model to be introduced in the present paper.

An obvious way to simplify the problem at hand is to assume spatial homogeneity
in the non-holographic spatial directions, so that the construction of the required finite density static solitons in the five-dimensional theory reduces to an effective one-dimensional problem.
However, it is easy to show \cite{Rozali:2007rx} that
there are no smooth spatially homogeneous gauge potentials that yield a non-zero baryon number density and have finite energy per baryon.
To circumvent this problem, the approach in \cite{Rozali:2007rx} 
imposes homogeneity through the use
of singular configurations. This reduces the problem to the determination
of a profile function that specifies the singluar gauge potential as a function of the holographic coordinate. An alternative way to obtain spatial
homogeneity is to replace the three-dimensional Euclidean spatial slice
by a three-sphere \cite{Kim:2007vd}. As there are smooth finite energy spatially
homogeneous gauge potentials on the three-sphere, this yields a finite
baryon number density that can be varied by changing the radius of the
three-sphere. However, in the studies in \cite{Kim:2007vd} the dependence on
the holographic direction is then assumed to be a fixed self-dual form, so
an effective one-dimensional kink model is not obtained and the possibility
of different types of behaviour in the holographic direction is excluded.

The approach taken in the present paper may be viewed as a hybrid of the above
two methods. We apply a homogeneous approximation at the level of the field strength, rather than at the level of the gauge potential, in a way that produces an expression for the field strength that is very similar to that obtained on the three-sphere. However, we
retain an arbitrary dependence on the holographic coordinate that leads
to the derivation of an effective kink theory in the holographic direction.
As the baryon chemical potential is increased
this kink theory yields a first order phase transition,
just below the baryon mass,
to lightly bound nuclear matter, as found in QCD.

We explain why our homogeneous approximation should be viewed as a
smeared crystal that is
expected to provide a lower bound on the energy, as it is an
unattainable idealization that distributes the energy perfectly equally in
space, which a true crystal is unable to match. We provide a justification for this expectation based on the simpler case of the Skyrme crystal.
Homogeneous approximations are often
employed to study field theory at very high density and
usually overestimate the true energy because of
the existence of symmetry breaking negative modes.
It is important to note that our smearing approximation
is in contrast to this standard application of homogeneity,
as smearing lowers the energy to provide an underestimate that is
expected to give valid predictions even for the small to moderate densities
at which bound nuclear matter is formed.

As baryon number density is increased it is natural to expect that the bulk
soliton explores more of the holographic direction and this has been
interpreted \cite{Kaplunovsky:2012gb} as a holographic realization of the
quarkyonic phase \cite{McLerran:2007qj}, where there is a quark Fermi sea
with a baryonic Fermi surface. It has also been proposed \cite{deBoer:2012ij}
that this provides a mechanism for approximate chiral symmetry restoration.
However, the details of the way in
which the bulk soliton expands into the holographic direction are unknown and
several different possibilities have been suggested.

One proposal is a
dyonic salt phase \cite{Rho:2009ym}, in which the bulk soliton splits into constituents at high density. This suggestion is motivated by a point particle
approximation and the analogy with calorons, which are flat space self-dual
instantons that can split into monopole constituents if a periodic direction
is smaller than the size of an instanton. We find that our effective kink model
provides an explicit realization of the dyonic salt phase, as the kink splits
into a pair of half-kink constituents as the density is increased.
Furthermore, there is a simple explanation for this splitting into constituents
in terms of the novel potential that appears in the effective kink model and
its explicit dependence on the holographic coordinate.

A second proposal, based on the use of approximations involving flat
space calorons and dilute instantons, is a baryonic popcorn phase
\cite{Kaplunovsky:2012gb,Kaplunovsky:2015zsa},
in which a sequence of transitions takes place
where the soliton crystal develops additional layers in the holographic
direction with increasing density. The kink model is also able to capture
the phenomenon of baryonic popcorn and displays a first order phase transition
that generates an additional soliton layer in the holographic direction
that is energetically preferred over the dyonic salt phase.
The fact that the phase transition of the first pop into baryonic popcorn occurs before the dyonic salt phase agrees with the results of full
numerical simulations of low-dimensional toy models
\cite{Bolognesi:2013jba,Elliot-Ripley:2015cma}, based on
sigma model instantons rather than Yang-Mills intantons.
However, we find that the kink model predicts only one pop, rather than the sequence of pops suggested by previous point particle
and flat space approximations \cite{Kaplunovsky:2012gb,Kaplunovsky:2015zsa}.
Once again, the
form of the potential that appears in the effective kink model provides
a simple explanation of the absence of additional pops once the first pop
has taken place.

In the kink model, the two layers produced by the single pop form the surface of a soliton bag that increases in size as the baryon chemical potential is increased. The interior of the bag is filled with abelian electric potential and the instanton charge density is localized on the surface of the bag. Our soliton bag is therefore
very similar to the magnetic bag \cite{Bolognesi:2005rk} that approximates a large number of coincident non-abelian BPS monopoles. A magnetic bag also has the
topological charge density localized on its surface, with the interior and the
exterior of the bag corresponding to different values of the modulus of the Higgs field. In our soliton bag the abelian electric potential plays the role of
the modulus of the Higgs field, taking different values inside and outside the bag. Note that our soliton bag is qualitatively different from the
previously proposed instanton bag \cite{Bolognesi:2014dja}, in which
monopole walls are embedded into the Sakai-Sugimoto model to produce
a bag where the interior is filled with instanton charge density, rather
than this density being localized on the surface of the bag.

\section{An effective holographic kink theory}\quad
In our study of the Sakai-Sugimoto model we choose to work
with the Yang-Mills theory that results from the
non-abelian Dirac-Born-Infeld action at leading order in $\alpha'$.
This simplification has the advantage that we do not need to choose a
particular proposal for the non-abelian form of the Dirac-Born-Infeld action.
The five-dimensional 
$U(2)$ Yang-Mills gauge theory version of the Sakai-Sugimoto model
involves a spacetime with a warped metric of the
form
\be
\label{metric}
ds^2 = H(z) \,dx_{\mu} dx^{\mu} +  \frac{1}{H(z)}dz^2,
\ee 
where $x_\mu,$ with $\mu=0,1,2,3,$ are the coordinates of four-dimensional
Minkowski spacetime and $z$ is the spatial coordinate in the additional
holographic direction. The warp factor in this expression is
$H(z)=(1+z^2)^{2/3}$
and the signature of the metric is $(-,+,+,+,+)$.

In this section we follow the notation of \cite{Bolognesi:2013nja},
in which the hermitian
$U(2)$ gauge potential is decomposed into a non-abelian $SU(2)$
component $A$ and an abelian $U(1)$ component $\hat A$, with lowercase latin
indices (excluding $z$) running over the three non-holographic spatial
directions $i=1,2,3$, whilst uppercase latin indices run over all
four spatial directions, $I=1,2,3,z.$
We immediately restrict to the case of time independent fields, where
the appropriate static ansatz is
\be
{A}_0=0, \qquad A_I={A}_I(x_{J}), \qquad  
\hat{A}_0=\hat{A}_0(x_{J}), \qquad \hat{A}_{I} = 0.
\ee
Then, in units of $N_c\lambda/(216\pi^3)$, where
$N_c$ is the number of colours and $\lambda$ is the 
't Hooft coupling,
the static energy of the five-dimensional Sakai-Sugimoto model is given by
\be
E=\int\bigg\{
-\frac{(\partial_i\hat A_0)^2}{2H^{1/2}}-\frac{H^{3/2}}{2}(\partial_z\hat A_0)^2
    +\frac{\mbox{Tr}(F_{ij}^2)}{2H^{1/2}}+H^{3/2}\mbox{Tr}(F_{iz}^2)
      +\frac{4}{\Lambda}\hat A_0\epsilon_{ijk}\mbox{Tr}(F_{iz}F_{jk})\bigg\}\,d^3x\,dz,
    \label{energy}  \ee
    where $\Lambda=8\lambda/(27\pi)$ is the rescaled 't Hooft coupling.
    The final term in the above expression arises from the Chern-Simons term.
    
Baryon number is identified with the $SU(2)$ instanton number of the soliton
      \be
      N=\frac{1}{8\pi^2}\int\epsilon_{ijk}\mbox{Tr}(F_{iz}F_{jk})\,d^3x\,dz,
      \label{charge}
      \ee
      and this provides the energy bound $E\ge 8\pi^2 |N|.$
      From the final term in (\ref{energy}) we see that the instanton charge
      density sources the abelian electric field.
      
      In the gauge $A_z=0$ the holonomy that carries the pion degrees of freedom
      \be
      U({\bf x})={\cal P}\mbox{exp}\bigg(i\int_{-\infty}^\infty A_z({\bf x},z)\,dz\bigg),
      \ee
      appears in the boundary condition for $A_i$, namely,
      \be
      A_i({\bf x},z)\to R_i({\bf x}) \quad \mbox{ as }\quad  z\to\infty, \qquad \mbox{ where } \quad R_i({\bf x})=i(\partial_iU)U^{-1}.
      \label{bc}
      \ee
      The holonomy $U({\bf x}):{\mathbb R}^3\to SU(2)$ carries the topology of the field configuration due to the equality $N=\mbox{deg}\,U\in\mathbb{Z}=\pi_3(SU(2))$, where the compactification of $\mathbb{R}^3$ is a consequence of the fact that $U({\bf x})$ tends to the identity matrix as $|{\bf x}|\to \infty.$ 
      Note that the pure gauge currents, $R_i({\bf x})$,
      satisfy the zero curvature condition
      \be
      \partial_iR_j-\partial_jR_i+i[R_i,R_j]=0.
      \label{zerocurvature}
      \ee
      The first step in deriving our effective kink model is to work
      in the gauge $A_z=0$ and apply the separable approximation
      \be A_i({\bf x},z)=R_i({\bf x})\psi(z),
      \label{sep}\ee
      where the real function $\psi(z)$ satisfies the boundary
      conditions $\psi(-\infty)=0$ and $\psi(\infty)=1$, in order to reproduce the correct boundary condition (\ref{bc}).
      Applying the separable approximation (\ref{sep}), and making use of the zero curvature relation, yields the gauge field strength
      \be
      F_{ij}=-i[R_i,R_j]\psi(1-\psi), \qquad F_{iz}=-R_i\psi'.
      \label{homog}
      \ee
      Our second step is to
      make a homogeneous approximation by assuming that the
      gauge field strength is independent of ${\bf x}$. Explicitly, we set
      $R_i=-\beta\sigma_i/2$, where
$\sigma_i$ denote the Pauli matrices and
      $\beta$ is a real constant that we take to be positive, as the
      sign of $\beta$ will turn out to be equal to the sign of the baryon
      number density.
            It is crucial that we apply the homogeneous approximation at the level of the field strength and not at the level of the gauge potential, since there are no homogeneous gauge potentials
that yield the formulae (\ref{homog}). This is because 
the zero curvature condition (\ref{zerocurvature}) is obviously
violated by restricting to constant non-commuting currents.
By imposing homogeneity of the
field strength, rather than the gauge potential, we have been able to
incorporate the zero curvature relation before restricting to homogeneous fields that violate it. If a homogeneous approximation is applied directly at the level of the gauge potential then the zero curvature relation (\ref{zerocurvature}) implies that the field is topologically trivial. The only way to reintroduce instanton charge is to incorporate it via a discontinuous gauge potential  \cite{Rozali:2007rx}, with the discontinuity being the source of the instanton charge.
Essentially, the winding of the field is then moved from infinity onto a discontinuity in the bulk. The advantage of our continuous approach is that we avoid the need to deal directly with the non-abelian gauge potential, by restricting attention to the continuous physical fields.

In summary, our homogeneous approximation is
\be
F_{ij}=\frac{1}{2}\beta^2\psi(1-\psi)\epsilon_{ijk}\sigma_k,
\qquad F_{iz}=\frac{1}{2}\beta \psi'\sigma_i, \qquad
\hat A_0=\omega(z).
\label{ansatz}
\ee
These expressions are very similar to those that appear if the spatial slice
$\mathbb{R}^3$ is replaced by $S^3$ with a finite radius,
as studied in \cite{Kim:2007vd}. In the case of the three-sphere,
smooth spatially homogeneous gauge potentials exist and 
directly generate relations analogous to (\ref{ansatz}), with
$\beta$ related to the inverse of the radius of the three-sphere.
However, in the work in \cite{Kim:2007vd} the analogue of
the kink function $\psi(z)$ is taken to be a fixed self-dual form,
so an effective kink model is not obtained. Taking a fixed form for the
kink function prevents a study of the way in which the bulk soliton
explores the holographic direction with increasing density
and the associated different phases.
This is the main purpose of the present paper. It might be interesting
to repeat the analysis presented here for the case of the
three-sphere with an unfrozen kink field, where genuine
homogeneous gauge potentials underpin the approximation.

To begin with, we work in the canonical ensemble with fixed baryon number density and no explicit baryon chemical potential. The boundary condition on the real function $\omega(z)$ is therefore $\omega(\pm\infty)=0,$
because in holographic QCD
the boundary value of $\hat A_0$ is proportional to the baryon chemical potential.

At finite density, the true spatial distribution of the fields in the non-holographic directions is expected to form a soliton crystal.
Our homogeneous approximation may be viewed as a smeared version of the crystal,
and we expect that this approximation  provides a lower bound on the true
crystal energy, as homogeneity is an unattainable idealization. A justification for this expectation is provided in the appendix, where we consider the related, though simpler, case of the Skyrme crystal. 

Substituting the expressions (\ref{ansatz}) into (\ref{energy}) and (\ref{charge}) gives the energy per unit 3-volume
\be
{\cal E}=\frac{1}{2}\int_{-\infty}^\infty\bigg\{
  3\beta^2H^{3/2}\psi'^2+\frac{3}{H^{1/2}}\beta^4\psi^2(1-\psi)^2-H^{3/2}\omega'^2
  +\frac{24}{\Lambda}\beta^3\omega\psi(1-\psi)\psi'\bigg\}\,dz
  \label{henergy}
  \ee
  and the baryon number density (ie. instanton number per unit 3-volume)
  \be
  \rho=\int_{-\infty}^\infty\frac{3\beta^3}{8\pi^2}\psi(1-\psi)\psi'\,dz
  =\frac{\beta^3}{16\pi^2}.
  \label{density}
  \ee
  The field equation for $\omega$ that follows from the variation of (\ref{henergy}) is
  \be
  \big(H^{3/2}\omega'\big)'=\frac{12\beta^3}{\Lambda}\bigg(\frac{1}{3}\psi^3-\frac{1}{2}\psi^2\bigg)'.
  \label{omegaeqn}
  \ee
  \begin{figure}[ht]
  \begin{center}
  \includegraphics[width=9cm]{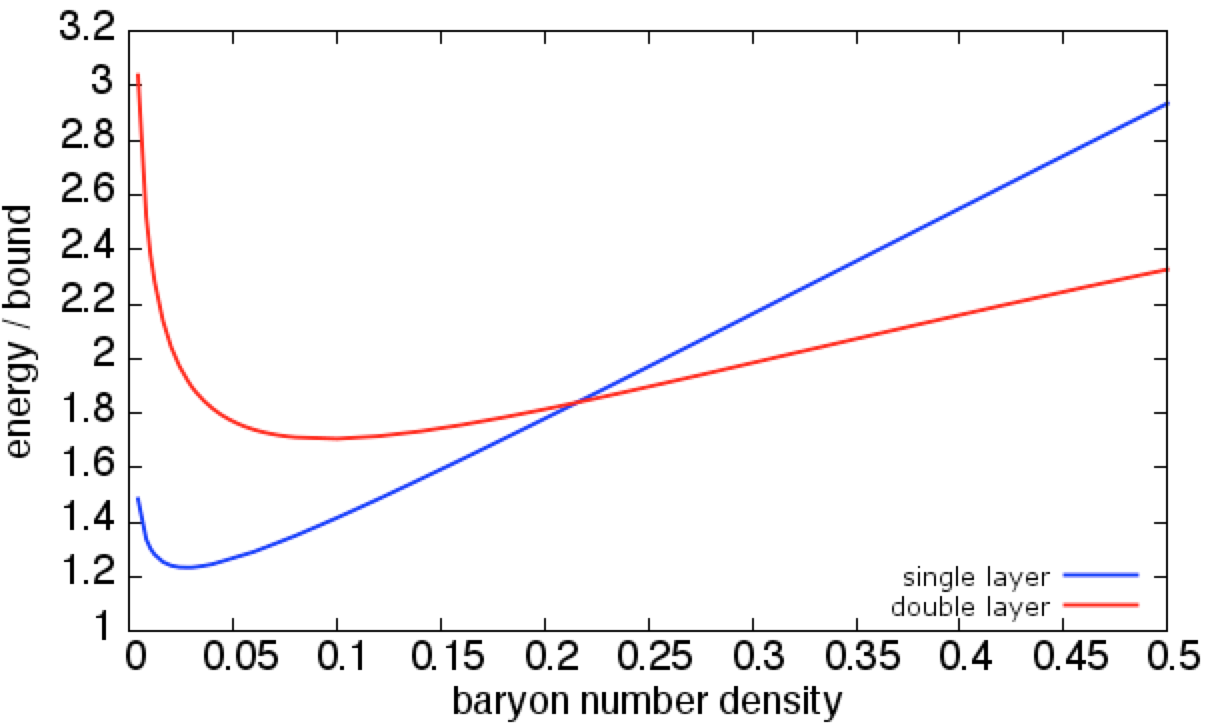}
    \caption{The ratio ${\cal E}/(8\pi^2\rho)$ of the energy to the energy bound as a function of baryon number density $\rho$. The blue curve is a single layer and the red curve is a double layer.}
\label{fig-energy}
  \end{center}
\end{figure}

As energy is minimized by a bulk configuration that is localized around $z=0$, we impose the symmetry that both the instanton charge density and $\omega$ are even functions of $z$, that is, $\omega(-z)=\omega(z)$ and $\psi(-z)=1-\psi(z)$. We can therefore restrict our discussion to the region
  $z\ge 0$, together
  with the boundary conditions $\omega'(0)=0$ and $\psi(0)=\frac{1}{2}$.
  Integrating (\ref{omegaeqn}) once and applying these boundary conditions at $z=0$ yields 
 \be
 H^{3/2}\omega'=\frac{12\beta^3}{\Lambda}\bigg(\frac{1}{3}\psi^3-\frac{1}{2}\psi^2+\frac{1}{12}\bigg).
 \label{wpp}
 \ee
  Taking the limit of this equation as $z\to\infty$ gives
  \be
  \lim_{z\to\infty}z^2\omega'=-\frac{\beta^3}{\Lambda}=-\frac{16\pi^2\rho}{\Lambda}.\ee
  We therefore obtain the usual holographic result, relating the coefficient
  of the asymptotic behaviour to the baryon number density,
  \be
  \omega=\frac{16\pi^2\rho}{\Lambda z}+o\bigg(\frac{1}{z}\bigg).
  \label{asymptotic}
  \ee
We can rewrite the final term in (\ref{henergy})     
 by applying an integration by parts to obtain the identity
  \be
  \int_{-\infty}^\infty\omega\psi(1-\psi)\psi'\,dz
  =\int_{-\infty}^\infty\omega'\bigg(\frac{1}{3}\psi^3-\frac{1}{2}\psi^2+\frac{1}{12}\bigg)\,dz.
  \ee
  Using this result, together with (\ref{wpp}), we obtain
  \be
     {\cal E}=\int_0^\infty\bigg\{
     3\beta^2H^{3/2}\psi'^2+\frac{3}{H^{1/2}}\beta^4\psi^2(1-\psi)^2
     +\frac{144\beta^6}{\Lambda^2H^{3/2}}\bigg(\frac{1}{3}\psi^3-\frac{1}{2}\psi^2+\frac{1}{12}\bigg)^2
     \bigg\}\,dz,
     \label{psionly}
  \ee
  where all reference to $\omega$ has been eliminated.
  This is the energy of our effective kink model.
 
  Note that imposing the flat space self-duality equation
$F_{IJ}=\frac{1}{2}\epsilon_{IJKL}F_{KL}$
  on the homogeneous fields (\ref{ansatz}) yields the first order equation
  \be\psi'=\beta\psi(1-\psi),\ee
  with kink solution
  \be
  \psi=\frac{1}{1+e^{-\beta z}}.
  \label{selfdual}
  \ee
  This is the self-dual form of the kink function that is assumed
  to hold for all densities in \cite{Kim:2007vd}.

  In the flat space limit $H=1$ with no Chern-Simons term ($\Lambda=\infty$), the energy (\ref{psionly}) of the self-dual solution (\ref{selfdual}) is
  ${\cal E}=\beta^3/2=8\pi^2\rho$, so the BPS behaviour of flat space instantons is recovered within our homogeneous approximation.
  
     \begin{figure}[ht]
  \begin{center}
    \includegraphics[width=9cm]{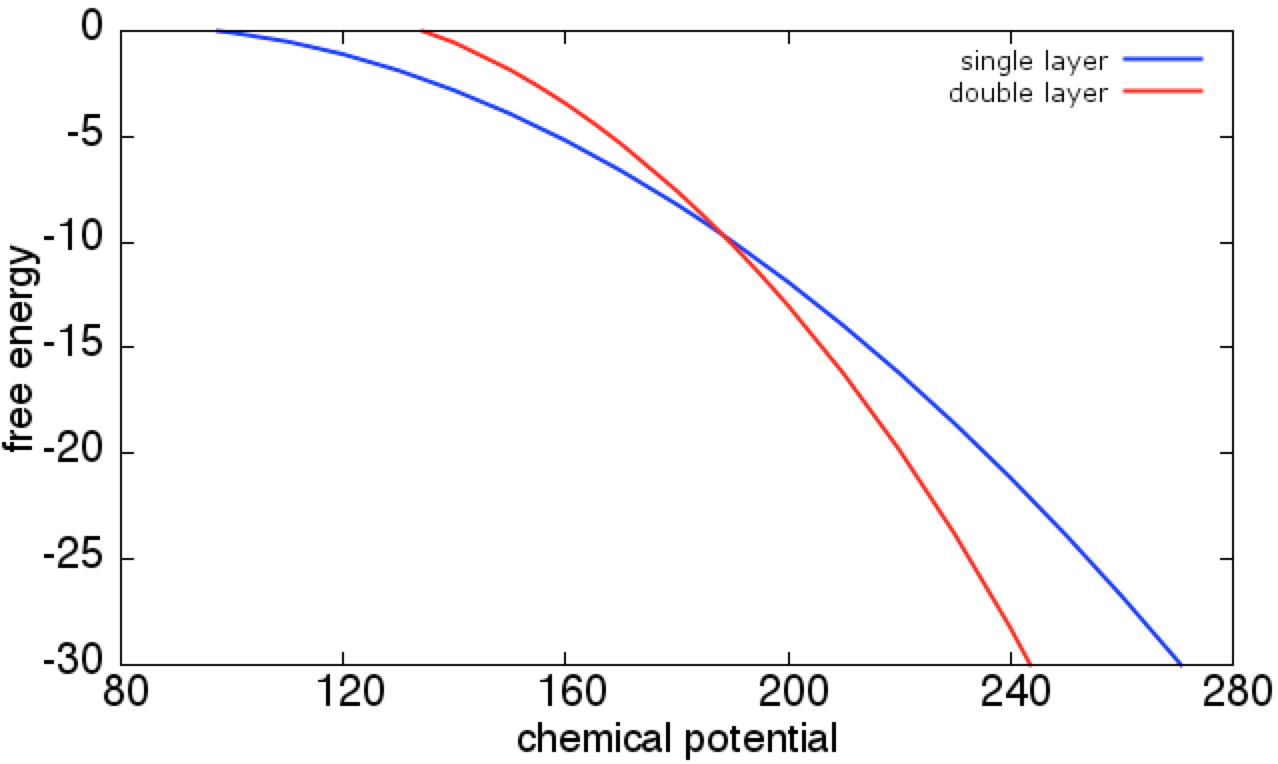}
    \caption{The free energy $\Omega$
      as a function of the chemical potential $\mu$ for a single layer (blue curve) and a double layer (red curve).}
\label{fig-freeenergy}
  \end{center}
     \end{figure}
\begin{figure}[ht]
  \begin{center}
  \includegraphics[width=9cm]{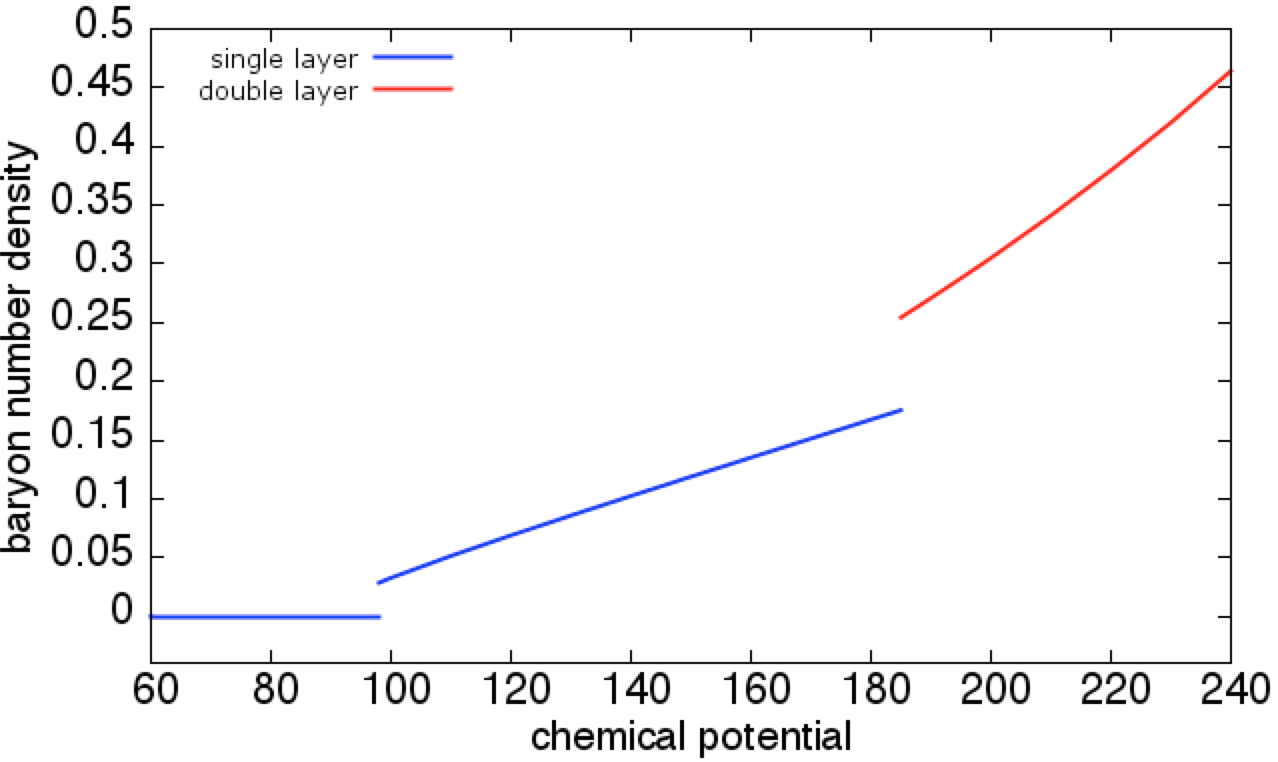}
    \caption{The baryon number density $\rho$ as a function of the chemical potential $\mu.$ The blue curve corresponds to a single layer and shows a first order phase transition just below the baryon mass. The red curve corresponds to a double layer and shows the first order phase transition of baryonic popcorn.}
\label{fig-chempot}
  \end{center}
\end{figure}
The field equation that follows from the variation of (\ref{psionly}) is
  \be
  \psi''+\frac{3H'}{2H}\psi'
  +\frac{\beta^2}{H^2}\psi(1-\psi)(2\psi-1)
  +\frac{48\beta^4}{\Lambda^2H^{3}}\psi(1-\psi)\bigg(\frac{1}{3}\psi^3-\frac{1}{2}\psi^2+\frac{1}{12}\bigg)=0.
  \label{psieqn}
  \ee
  We solve this equation numerically using a gradient flow algorithm and the
  change of variable $z=\tan u,$ to map the infinite range of $z$ onto a finite range of $u.$ All the numerical results presented in this paper are computed with the fiducial value $\Lambda=10.$

  In Figure~\ref{fig-energy} the blue curve shows a plot of the ratio of the energy to the lower bound, that is ${\cal E}/(8\pi^2\rho)$, as a function of the baryon number density $\rho$. We see that there is a non-zero optimum density
  $\rho_\star= 0.027$, 
  corresponding to the analogue of the nuclear matter density in QCD.
  As described below, this optimal density is associated with a critical value of the baryon chemical potential at which there is a first order phase transition to an equilibrium density of baryons.
\begin{figure}[ht]
  \begin{center}
    \includegraphics[width=6.5cm]{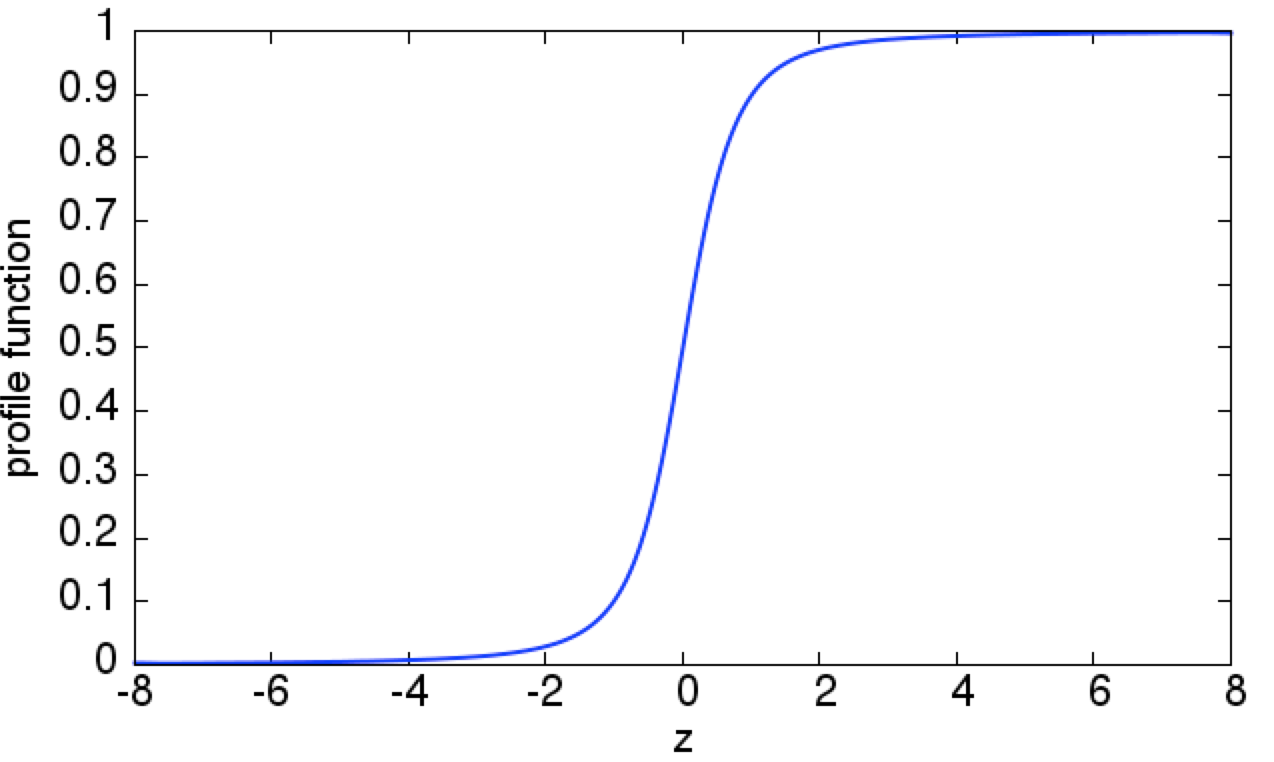}  \includegraphics[width=6.5cm]{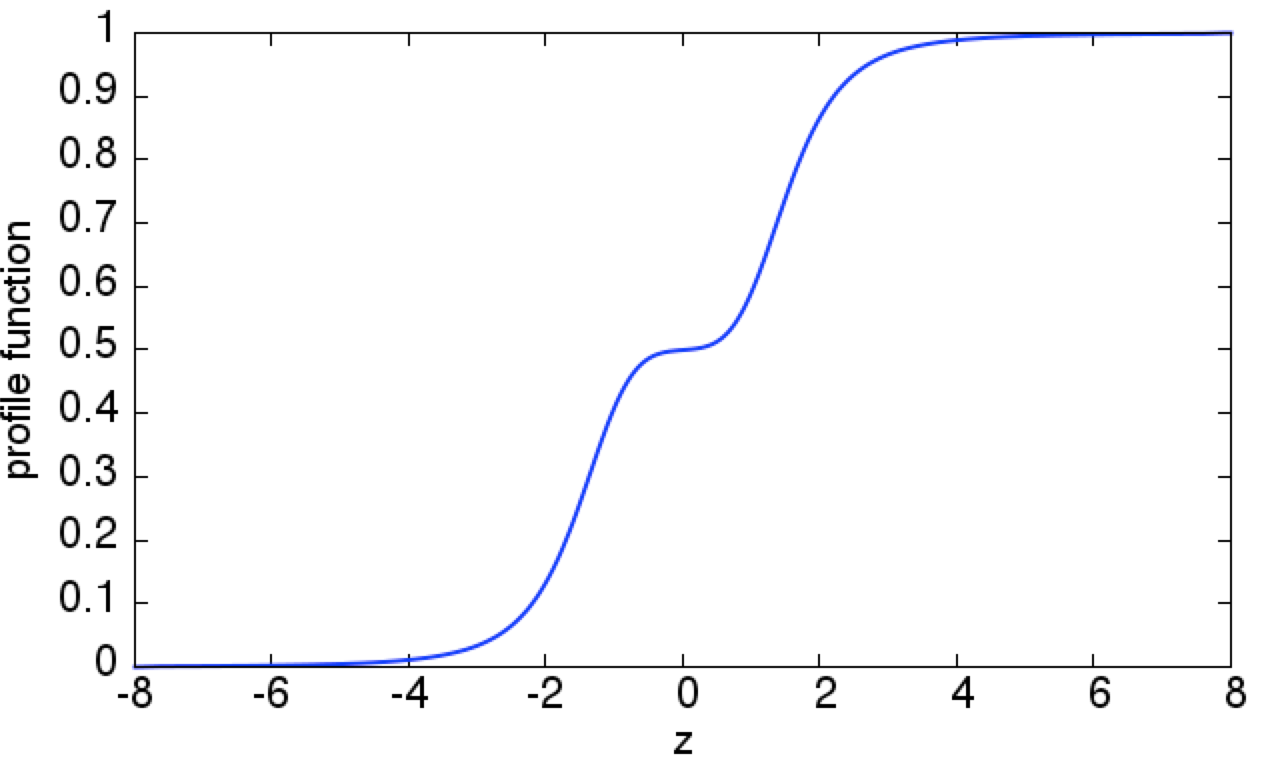}
    \includegraphics[width=6.5cm]{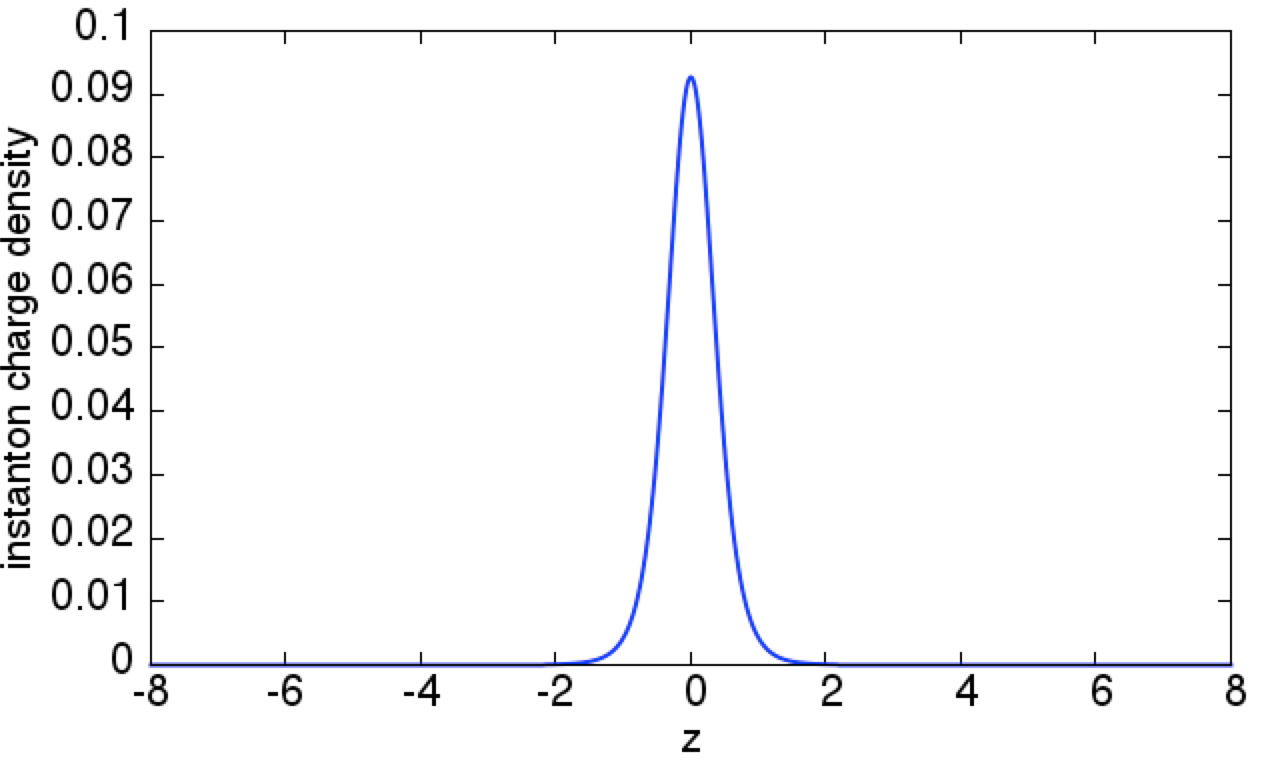}  \includegraphics[width=6.5cm]{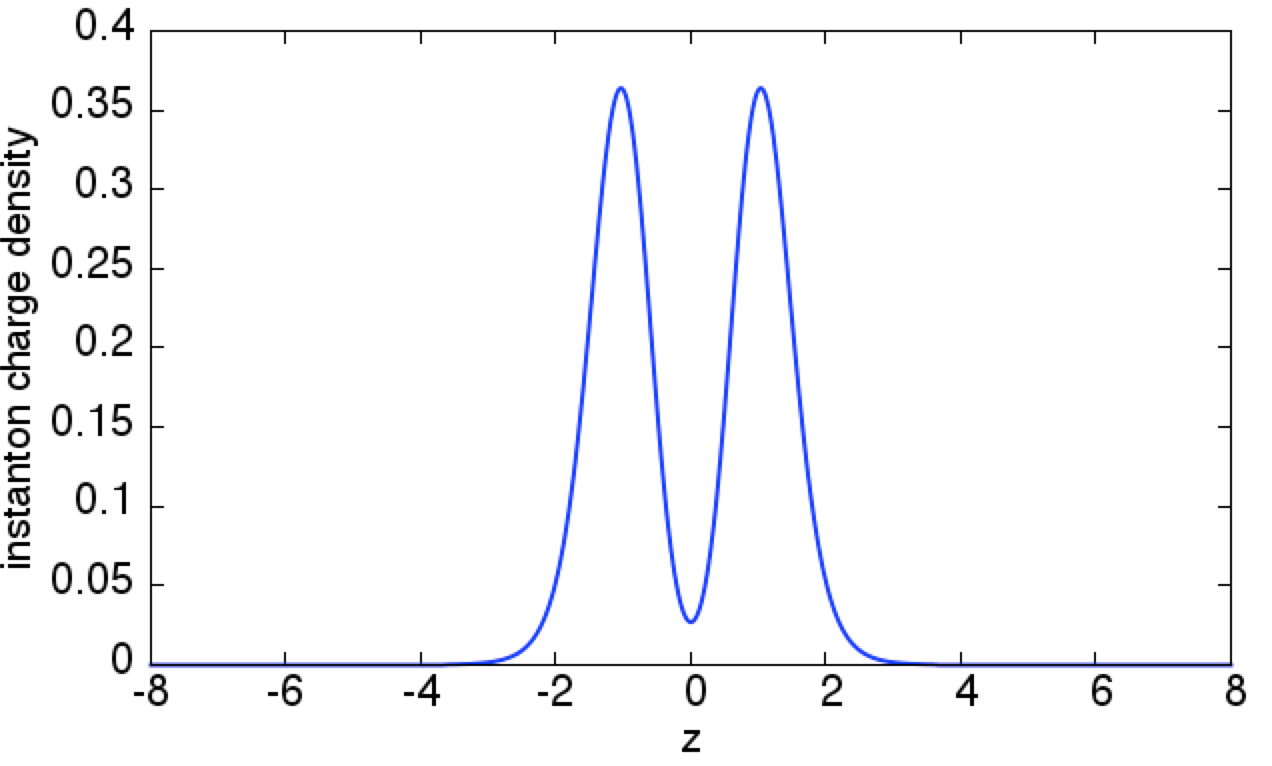}
    \includegraphics[width=6.5cm]{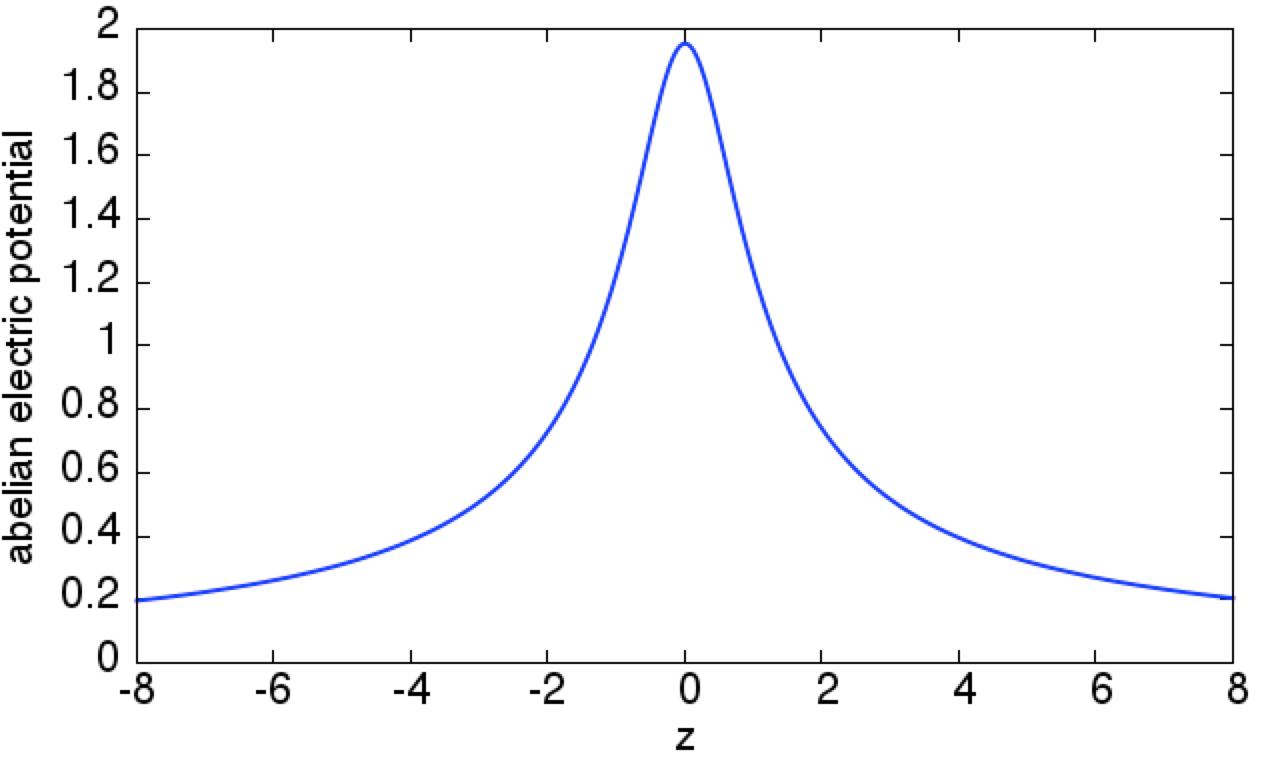}  \includegraphics[width=6.5cm]{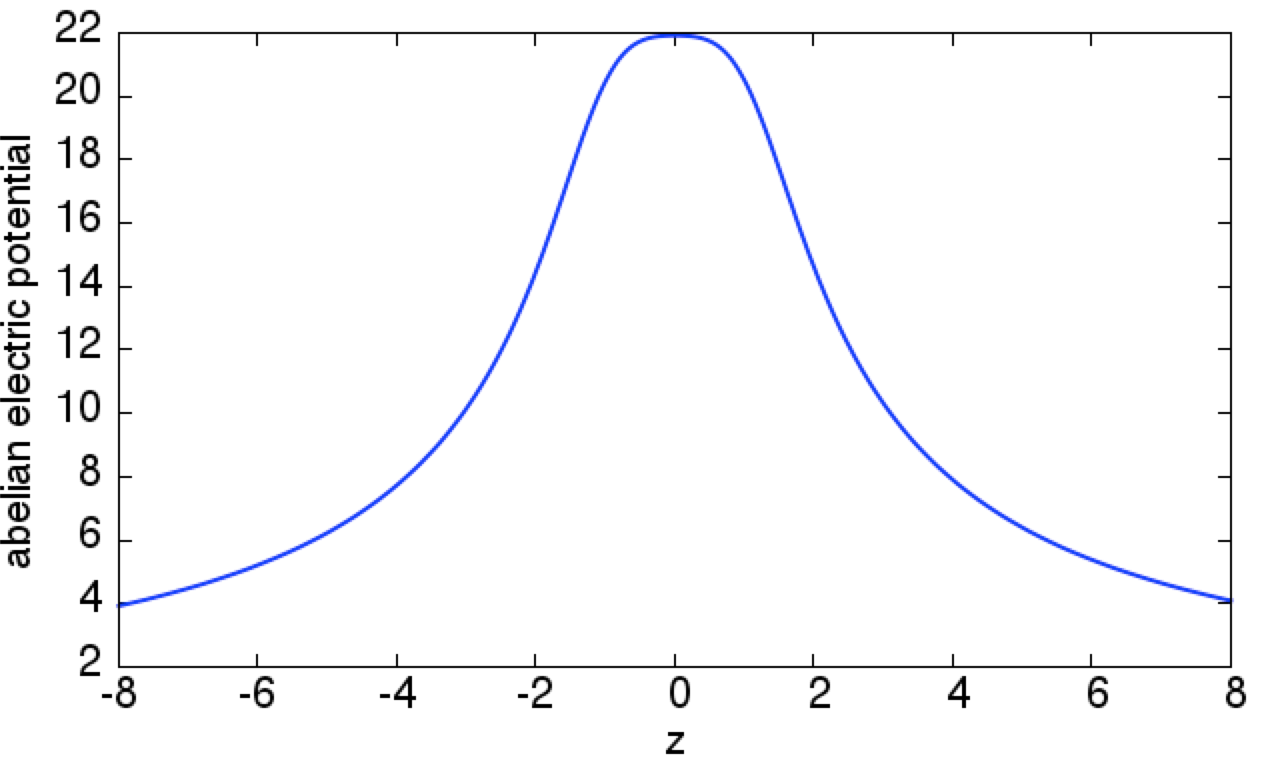}
  \caption{The single layer profile function $\psi(z)$ (top row), the instanton charge density per unit 3-volume (middle row) and the abelian electric potential $\omega(z)$ (bottom row) for baryon number densities $\rho=0.1$ (left column) and $\rho=2$ (right column). At high density the kink splits into a pair of half-kink constituents.}
\label{fig-psi}
  \end{center}
\end{figure}

  In the above we have worked in the canonical ensemble with fixed baryon number density, but an alternative is to work in the grand canonical ensemble with
  fixed baryon chemical potential $\mu$. Applying the standard holographic dictionary, in the grand canonical ensemble a baryon chemical potential corresponds to a non-zero boundary value for $\hat A_0$.  With our chosen normalizations, a vectorial chemical potential $\mu$ corresponds to the boundary condition
\be
\omega(\pm\infty)= -\frac{\mu \Lambda}{32\pi^2}.
\label{chempot}
\ee
This is because a shift in $\omega$ by this constant value returns us to the previous boundary condition $\omega(\pm\infty)=0$, but from (\ref{henergy}) we see that the energy then transforms into the grand potential
(also known as the Landau free energy) $\Omega={\cal E}-\mu\rho.$ Minimizing $\Omega$ at fixed $\mu$ we obtain the function $\Omega(\mu)$ displayed as the blue curve in Figure~\ref{fig-freeenergy}. The associated
relation between the chemical potential $\mu$ and the baryon number density $\rho$ is displayed as the blue curve in Figure~\ref{fig-chempot}. These curves show
that a non-zero value for $\rho$ is obtained for $\mu\ge\mu_\star,$ where $\mu_\star=98$ (for $\Lambda=10$) is the critical value of the chemical potential at which the density is indeed equal to $\rho_\star$.

For attraction between baryons, the critical value of the chemical potential $\mu_\star$ must be less than the baryon mass $M_B.$ In the units we are using, the self-dual single instanton approximation to the baryon yields the formula
 \cite{Bolognesi:2013nja}
\be
M_B=2\pi^2\bigg(4+\frac{32}{\Lambda}\sqrt{\frac{2}{15}}\bigg),
\label{baryonmass}
\ee
where terms of order $1/\Lambda^2$ have been neglected. Substituting our fiducial value $\Lambda=10$ into (\ref{baryonmass}) gives $M_B=102,$
so indeed our numerically computed value $\mu_\star=98$ is less then $M_B$ and baryons form bound states. The percentage binding energy per nucleon is given by
\be
\Delta = (M_B-\mu_\star)/M_B\times 100\%,
\ee
yielding $\Delta=4\%$ for the chosen value $\Lambda=10.$
In the limit $\Lambda\to\infty$ the BPS result is recovered, $M_B\to 8\pi^2$ and $\mu_\star\to M_B$, giving zero binding energy.
Although we expect holographic QCD to provide only qualitative information about QCD with three colours, the above results imply that there must be a value of $\Lambda$ greater than 10 at which a realistic nuclear binding energy
$\Delta \sim 0.9\%$ is obtained, and indeed we compute that
the appropriate value is $\Lambda\sim 18$.   

Returning to the representative value $\Lambda=10$,
in the top row in Figure~\ref{fig-psi} we display the kink profile function $\psi(z)$ for the densities $\rho=0.1$ (left column) and $\rho=2$ (right column).
In the middle row
we plot the corresponding instanton number densities per unit 3-volume, and in
the final row we display the abelian electric potential $\omega(z)$.
We see that at high density the kink splits into two half-kink constituents. This splitting of the soliton layer into a pair of constituents is the predicted dyonic salt phenomenon, corresponding to the fact that a periodic instanton (a caloron) splits into monopole constituents at high density.
Note that a half-kink cannot exist in isolation as a finite energy configuration because the profile function $\psi(z)$ of a half-kink interpolates between values that differ by $\frac{1}{2}$, but for finite energy this difference must be equal to $\pm1$ or $0$.  This mirrors the caloron situation, where the caloron splits into monopole constituents that are not finite energy configurations in isolation but together combine to form a well-defined periodic instanton.

The form of the energy (\ref{psionly})
of the effective kink model provides a simple
explanation for the split into half-kink constituents, as follows.
The kink model contains a derivative term plus two potential terms, which
are novel in a kink theory due to the form of the explicit
dependence on the holographic spatial coordinate.
The first potential term has an explicit spatial dependence that grows like
$z^{-2/3}$ for large $z$. Such a term does not decay fast enough to be
integrable, so the associated multiplying factor must tend to zero
as $z\to\pm\infty.$ This enforces the boundary conditions 
$\psi(\pm\infty)\in\{0,1\}.$
The second potential term has an explicit spatial dependence that grows like
$z^{-2}$, which decays fast enough that finite energy considerations impose
no conditions on the boundary values of $\psi(z)$ from this term.
However, as this term has a coefficient of $\beta^6$ then at high density
it is, at least locally, of more relevance than the first potential term,
which has a coefficient of only $\beta^4$. The second potential term
is minimized by the value $\psi=\frac{1}{2}$, as it vanishes at this value.
Putting all this toegther we see that as the density increases
the second potential term induces the kink field $\psi(z)$ to remain close to
the value $\frac{1}{2}$ over an increasingly large range of $z$, although
the finite energy requirement of the first potential term always forces
the kink boundary conditions to ultimately be attained. This is the
simple reason why the kink splits into a pair of half-kinks.

Previous results on low-dimensional models 
\cite{Bolognesi:2013jba,Elliot-Ripley:2015cma}
suggest that a baryonic popcorn transition 
\cite{Kaplunovsky:2012gb}
appears before the formation of dyonic salt. In other words, 
at densities high enough to split the kink into a pair of half-kink constituents, we expect that there is a competing solution that describes a double
layer and has lower energy than the single layer considered in this section.
In the following section we examine this possibility by 
constructing a double layer configuration and calculating its energy as a
function of baryon number density. 

\begin{figure}[ht]
  \begin{center}
    \includegraphics[width=6.5cm]{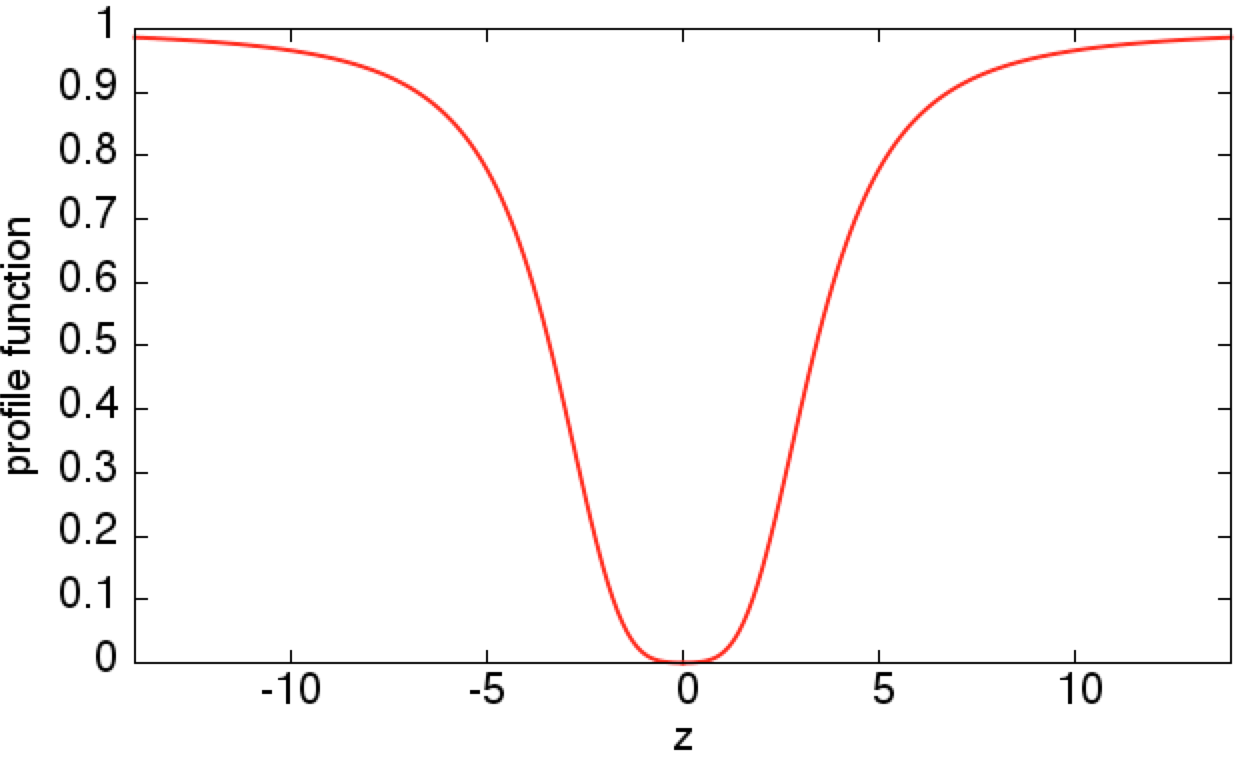}  \includegraphics[width=6.5cm]{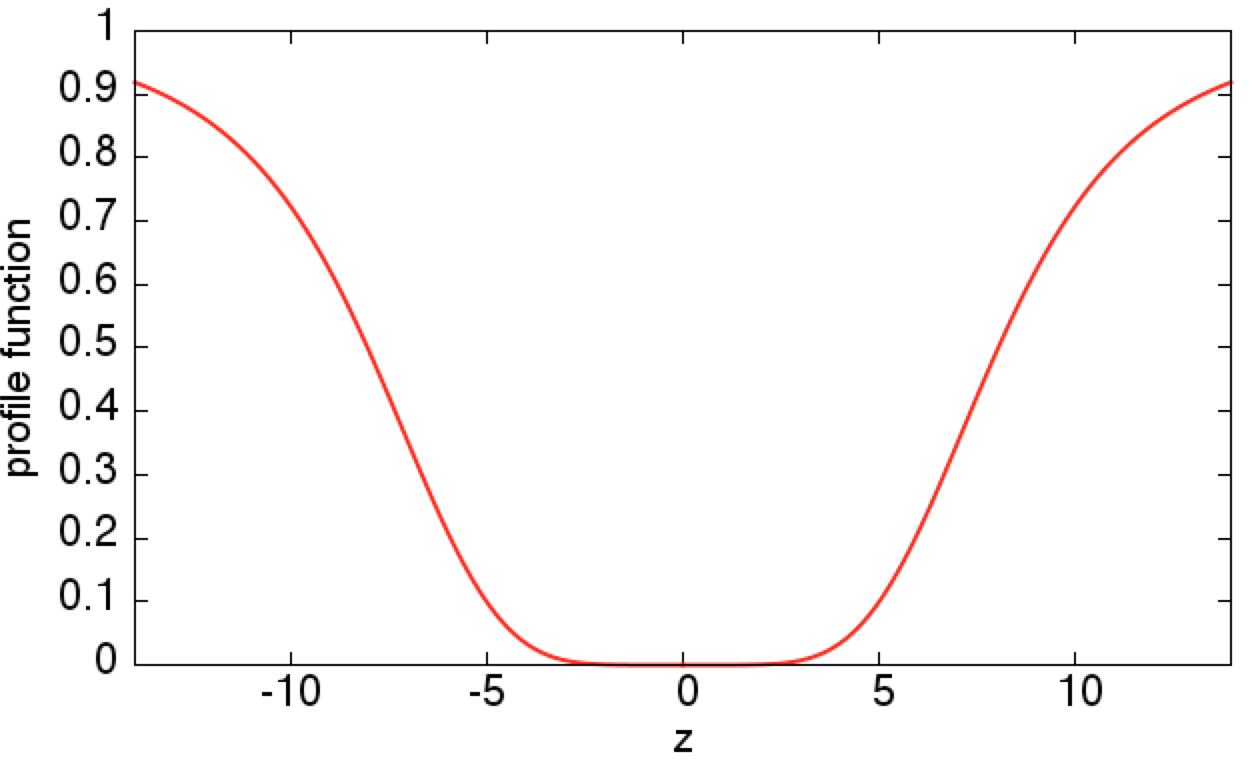}
    \includegraphics[width=6.5cm]{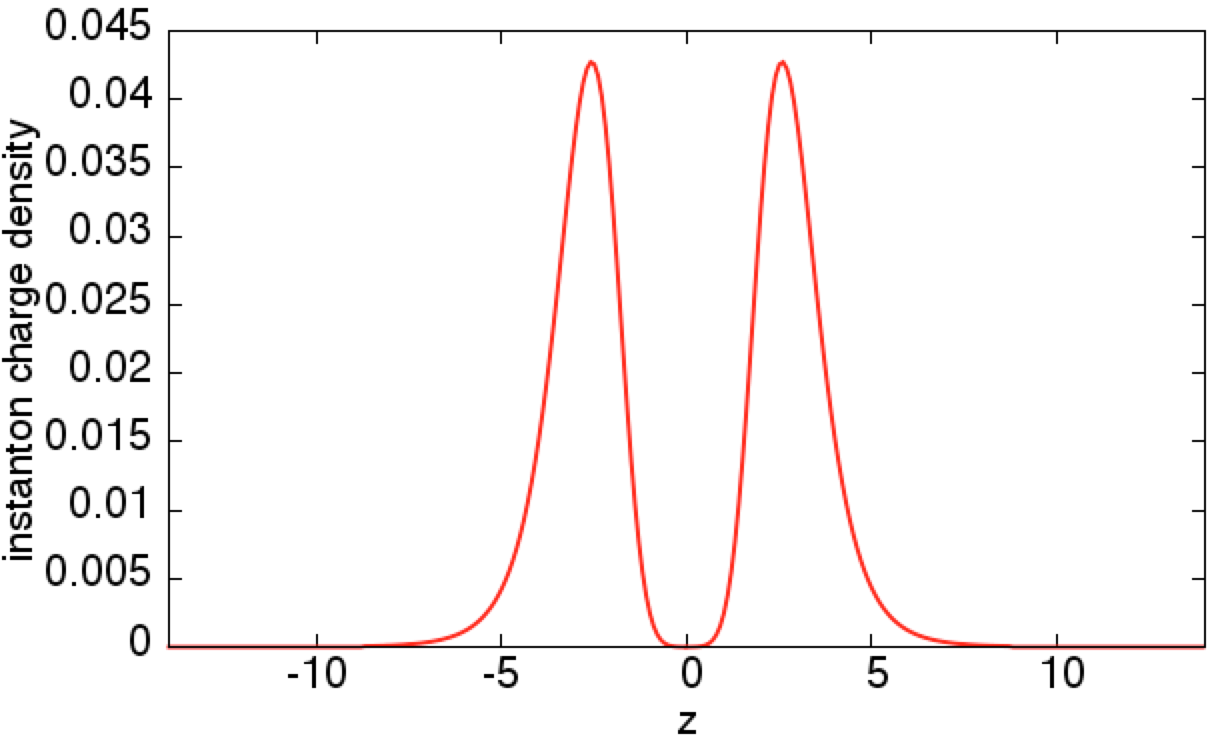}  \includegraphics[width=6.5cm]{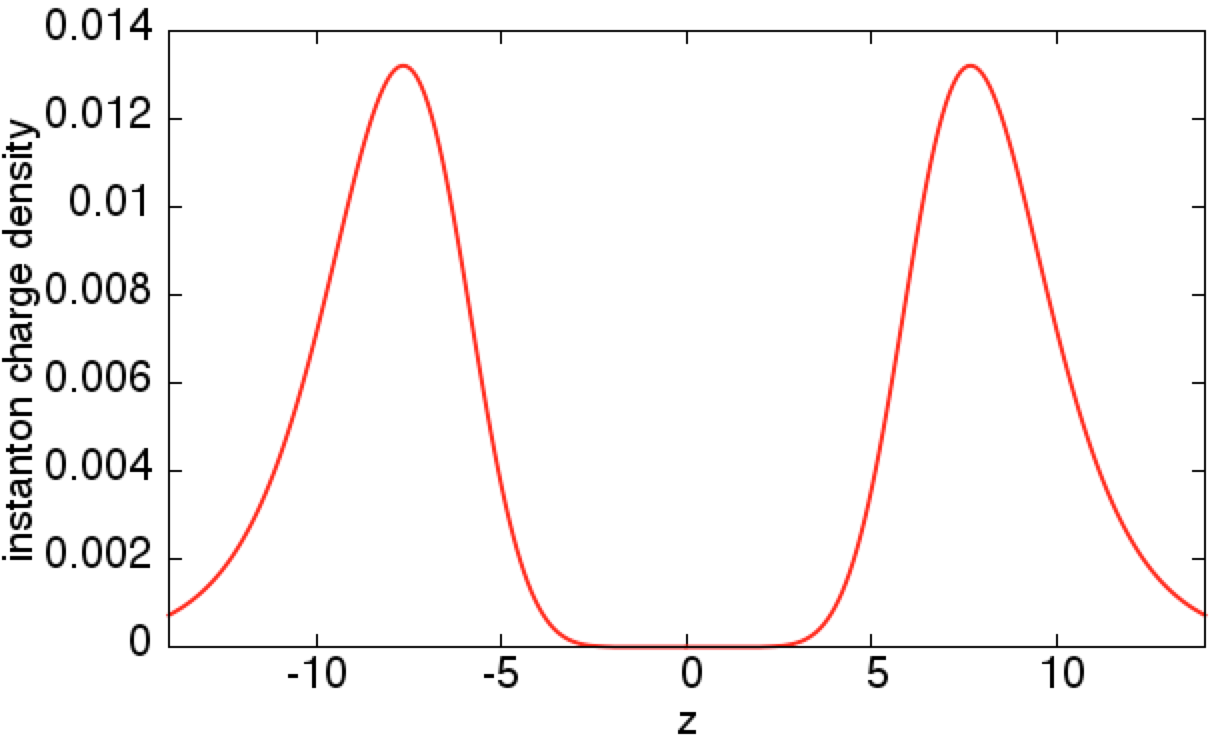}
    \includegraphics[width=6.5cm]{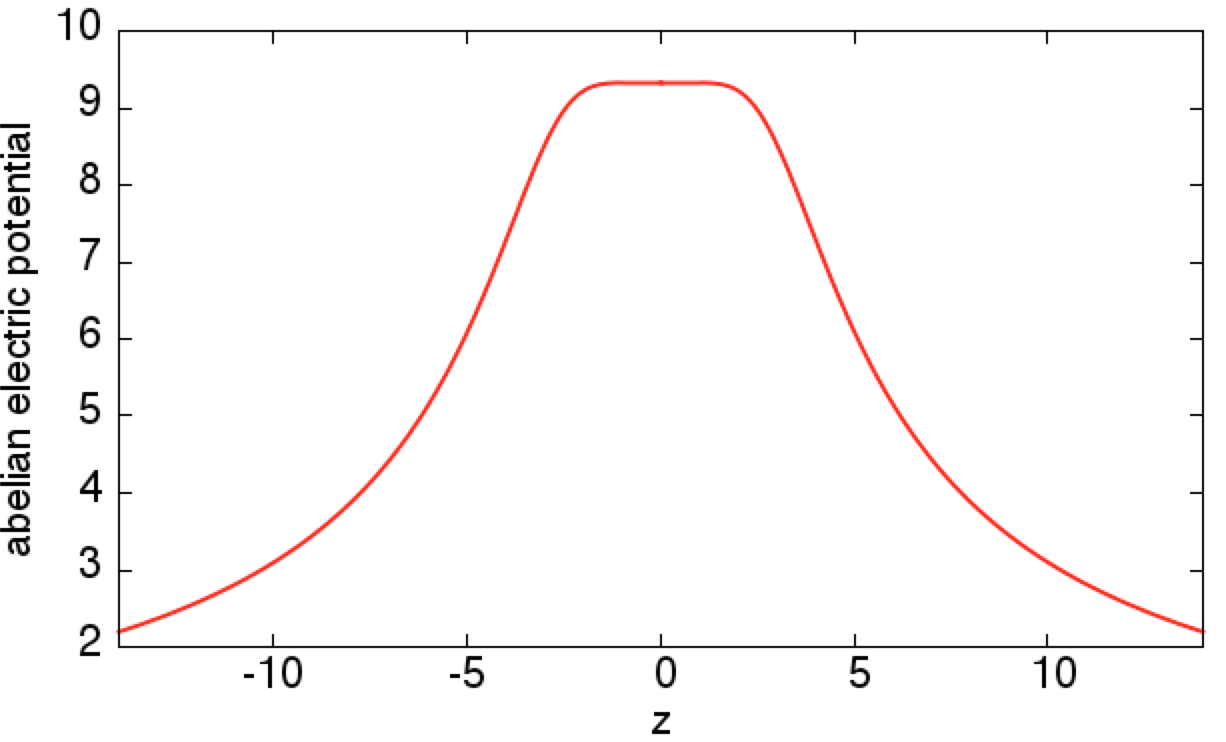}  \includegraphics[width=6.5cm]{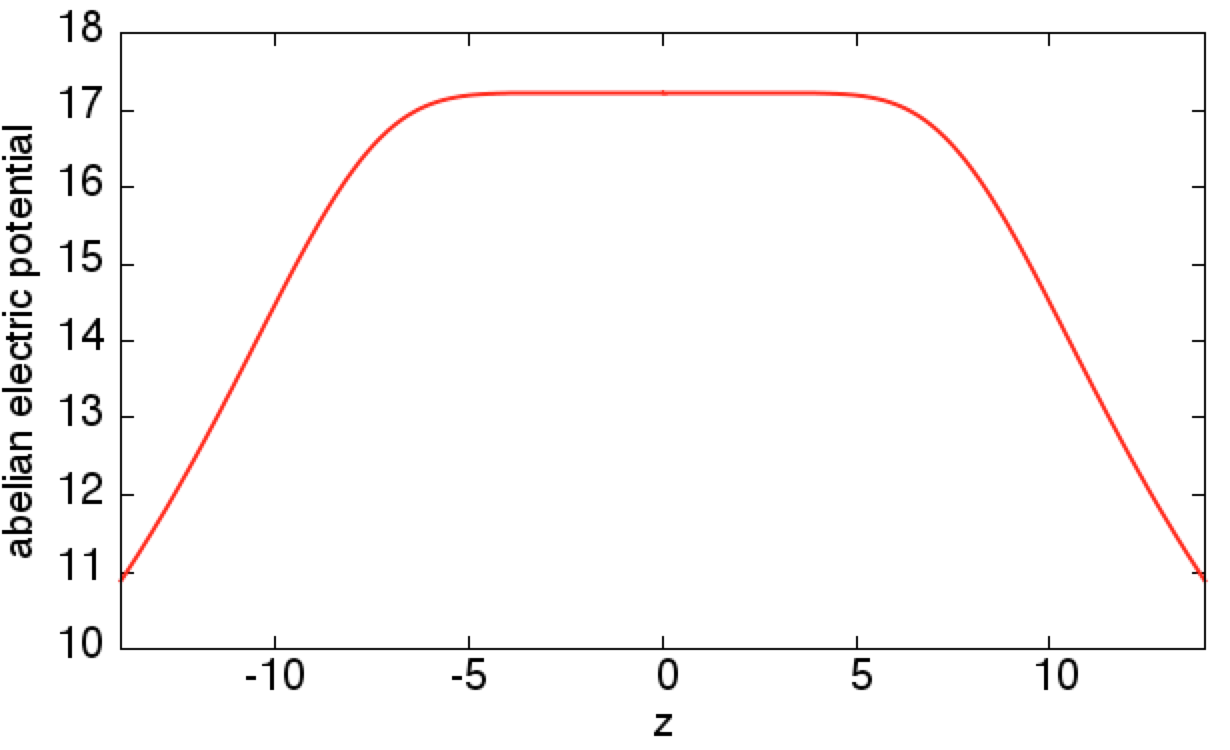}
  \caption{The double layer profile function $\psi(z)$ (top row), the instanton charge density per unit 3-volume (middle row) and the abelian electric potential  $\omega(z)$ (bottom row) for baryon number densities $\rho=2$ (left column) and $\rho=10$ (right column).}
\label{fig-psi2}
  \end{center}
\end{figure}

\section{Baryonic popcorn and soliton bags}\quad
The effective kink model, with energy given by (\ref{psionly}), has an anti-kink
solution with boundary conditions $\psi(-\infty)=1$ and $\psi(\infty)=0.$ It is obtained from the kink solution by the transformation $\psi\mapsto 1-\psi$ and has the same energy per unit volume as the kink but has a negative baryon number density. However, if the ansatz (\ref{ansatz}) is modified by the replacement
$F_{ij}\mapsto -F_{ij}$ then the anti-kink now has a positive baryon number
density, and this provides an equivalent anti-kink formulation of the single
layer described in the previous section in terms of a kink.

We can construct a double layer configuration by gluing together two single layers in a continuous manner. Explicitly, we apply the approximation (\ref{ansatz}) for $z\ge 0$ with kink boundary conditions on the half-line, $\psi(0)=0$ and $\psi(\infty)=1.$ For $z\le 0$ we take the same approximation (\ref{ansatz}) after the replacement $F_{ij}\to -F_{ij}$ with anti-kink boundary conditions
on this half-line, $\psi(-\infty)=1$ and $\psi(0)=0.$
This corresponds to taking both
$\psi(z)$ and $\omega(z)$ to be even functions of $z$ and
gives a continuous instanton charge density.
The instanton charge density vanishes at
the join at $z=0$, and although it is continuous at
this point it is not smooth there.
Note that the parity of $\omega$ is as before, in agreement with the
requirement that we consider a vectorial baryon chemical potential.

The baryon number density for this two layer configuration is
\be
  \rho=\int_{0}^\infty\frac{6\beta^3}{8\pi^2}\psi(1-\psi)\psi'\,dz
  =\frac{\beta^3}{8\pi^2},
  \ee
  which is twice the value (\ref{density}) for the single layer.
  Integrating the field equation (\ref{omegaeqn}) for $\omega$,
  and imposing the new boundary condition $\psi(0)=0$ and
  $\omega'(0)=0$, gives for $z\ge 0$
  \be
 H^{3/2}\omega'=\frac{12\beta^3}{\Lambda}\bigg(\frac{1}{3}\psi^3-\frac{1}{2}\psi^2\bigg).
 \label{wpp2}
 \ee
  Taking the limit of this equation as $z\to\infty$ provides the relation
  \be
  \lim_{z\to\infty}z^2\omega'=-\frac{2\beta^3}{\Lambda}=-\frac{16\pi^2\rho}{\Lambda},\ee
  again reproducing the correct asymptotic behaviour (\ref{asymptotic}).

  As in the single layer case, an integration by parts yields the following expression for the double layer energy, 
    \be
     {\cal E}=\int_0^\infty\bigg\{
     3\beta^2H^{3/2}\psi'^2+\frac{3}{H^{1/2}}\beta^4\psi^2(1-\psi)^2
     +\frac{144\beta^6}{\Lambda^2H^{3/2}}\bigg(\frac{1}{3}\psi^3-\frac{1}{2}\psi^2\bigg)^2
     \bigg\}\,dz,
     \label{psionly2}
     \ee
     as a function of $\psi$ only. We obtain the double layer solution by
     numerically solving the static field equation for $\psi(z)$
     that follows from the variation of this energy.
     The red curve in Figure~\ref{fig-energy} is a plot of the ratio of the energy to the lower bound, as a function of the baryon number density, for
     the double layer solution. From this figure we see that the double layer has a
     lower energy than the single layer beyond the critical density $\rho_2= 0.22.$
     This is the critical density for a homogeneous baryonic popcorn transition
     to a double layer, and is well below the density at which dyonic salt
     appears. This result is therefore in agreement with the previous
     low-dimensional studies mentioned earlier, where a popcorn transition
     also appears before the dyonic salt phase.

     In the top row in Figure~\ref{fig-psi2}
     we display the double layer profile function $\psi(z)$ for the densities $\rho=2$ (left column) and $\rho=10$ (right column).
In the middle row
we plot the corresponding instanton number densities per unit 3-volume, and in
the final row we display the abelian electric potential $\omega(z)$.
Although there are some qualitative similarities between a double layer
and a single layer that has split into two half-layer constituents,
the crucial distinction is that a half-layer cannot exist in isolation
as a finite energy configuration. There are, of course, significant
quantitative differences, including the fact that the double layer
configuration has a much lower energy at high densities.

The free energy of the double layer as a function of the chemical potential is shown as the red curve in Figure~\ref{fig-freeenergy}. This reveals that within the grand canonical ensemble there is a critical value of the chemical potential,
given by
$\mu\ge\mu_2=185$ for $\Lambda=10$,
above which the free energy of the double layer is less than that of a single layer. This results in the baryonic popcorn first order phase transition seen in Figure~\ref{fig-chempot} at $\mu=\mu_2$, where the configuration pops from the single layer (given by the blue curve) to the double layer (given by the red curve).
Note that in Figure 3 we only plot the single layer and double layer portions
of the curve for the range of chemical potentials at
which each is the global minimum of the free energy.
The curves extend beyond these segments to
physically irrelevant regions in which they are no
longer global minima.

Given the results in the previous section, where a single layer splits into
half-layer constituents at high density, one might naively expect a
similar phenomenon to take place for a double layer, with each layer
splitting into half-layer constituents at high density.
However, the right column in Figure~\ref{fig-psi2} demonstrates
that the double layer does not split, even at very high densities.
This has a simple explanation, again obtained by examining the
form of the novel potential terms in the effective energy (\ref{psionly2}).
This time, we see that the final potential term in this effective energy
is minimal only at $\psi=0$ (for $\psi\in[0,1]$). Therefore, unlike the
case of a single kink, the final potential term does not induce the
kink or anti-kink to stay close to any new value as the density is increased.
Hence the double layer does not split, but rather the two layers
simply increase their separation as the density increases.

As the density increases and the two layers move further apart, it would
be reasonable to surmise
that further pops would occur, since there now appears to be available
space around $z=0$ to generate new layers. However, this simple view
ignores the fact that the region between the two layers is not empty but
instead is full of abelian electric potential, as demonstrated by the plots
in the bottom row of Figure~\ref{fig-psi2}. The two layers form the surface
of a soliton bag, with the interior of the bag associated with an approximately
constant non-zero value of the abelian electric potential. The abelian electric
potential decays to zero outside the bag, with the surface of the bag being
the transition
region where both the instanton charge density and the electric field
are localized. This is similar to the magnetic bag 
description of large charge non-abelian monopoles \cite{Bolognesi:2005rk},
where the surface of the bag separates regions of zero and non-zero
values for the modulus of the Higgs field.
As the system under consideration is periodic
in three spatial directions (approximated by homogeneity) then the
surface of the bag is not a single 
connected component, like the magnetic bag in three-dimensional
Euclidean space, but instead consists of
two disconnected components corresponding to
the top and bottom of the bag. This is why the soliton bag requires two layers.

The creation of more layers through further baryonic popcorn transitions
would produce a kind of multi-layer bag. In the monopole context,
the possibility of multi-layer magnetic bags has been investigated
in \cite{Manton:2011vm} with the conclusion that these typically consist
of only a single bag surrounded by layers of isolated unit charge monopoles.
Any attempt to create a multi-bag configuration automatically
rules out an interior bag that carries any significant fraction of the
total magnetic charge.
The soliton bag description therefore suggests the absence of additional
baryonic popcorn transitions. Further support for this view
is obtained by extending our anti-kink plus kink double layer approximation,
denoted $\bar K K$,
to additional layers by the inclusion of more anti-kinks/kinks.
For example, a 4-layer $\bar K K\bar K K$ approximation has been studied
where all contiguous anti-kinks and kinks are joined in the same continuous
manner as in the double layer approximation. The location of the joins,
together with the fraction of instanton charge carried by each layer, are
allowed to vary and the resulting energy minimizing configurations
computed. As expected from the similar monopole magnetic bag story, these
computations yield only signficiant instanton charge density in the
outer layer and the inner layer is irrelevant.
A recent study \cite{Preis:2016fsp}, applying a different approximation,
also reached the same conclusion that a double layer is the
preferred configuration at high density and
additional layers are not generated.

Finally, it is important to note that our
soliton bag is qualitatively different from the instanton bag
proposed in \cite{Bolognesi:2014dja} as a description of the high
density phase in the Sakai-Sugimoto model.
The instanton bag is obtained via an initial compactification of one of the
spatial directions to allow the embedding of a monopole wall.
A pair of monopole walls are then patched together to form the
surface of the instanton bag, with the result that the interior is
filled with instanton charge density. This contrasts with our soliton bag,
where the instanton charge density is localized on the surface of the bag.
Our kink approximation, with an appropriate shape for the kink profile
function, could produce a bag filled with instanton charge density, but
such a shape does not appear when the profile function is obtained by
minimizing the energy of the effective kink model.
      
\section{Conclusion}\quad
It is an open problem to understand the phases of cold and dense holographic
nuclear matter as a function of baryon number density. The distribution
of baryonic matter in both the non-holographic and
holographic directions is
unknown and this has led to a number of different approximate descriptions
and suggestions for phenomena that might occur. In this paper we have
assumed homogeneity in the non-holographic spatial directions to
investigate the distribution in the holographic direction. As the
holographic coordinate
corresponds to an energy scale in the boundary theory, then understanding this
aspect is likely to be the key to a holographic description of the
baryonic Fermi surface of a quarkyonic phase.

Our homogeneous holographic nuclear matter is described by an effective
holographic kink theory, which we have shown is capable of a
simultaneous realization of a number of previously suggested phases,
including dyonic salt, baryonic popcorn and soliton bags.
An advantage of this unifying description is that we can compare the
various phases and determine which phase is preferred as the
baryon chemical potential varies. The effective kink model also
successfully reproduces the QCD behaviour of a first order phase
transition to lightly bound nuclear matter, 
at a value of the baryon chemical potential that is
just below the baryon mass, in agreement with QCD.

In holographic QCD the spectrum of fluctuations in the holographic
direction determines the masses of the vector and axial vector mesons.
As the background solution changes with increasing baryon number density
then so does the spectrum, and this has been proposed \cite{deBoer:2012ij}
as a mechanism for approximate chiral symmetry restoration, as the
masses of vector and axial vector mesons could converge with increasing
density.
However, as the details of the background solution are unavailable,
to date the investigations of this issue have been mostly qualitative.
It would be interesting to study this aspect within our effective kink
model and to compute some quantitative data. Even within
our homogeneous approximation, this is still a non-trivial
calculation, because of the mixing of various modes, but it should be
a tractable problem that we hope to address in the near future.

Other avenues for future research include extending our approach to the
finite temperature regime and investigating how kinky
holographic nuclear matter responds to external electric and magnetic
fields by introducing new boundary conditions for $\hat A_\mu$, as
in \cite{Bergman:2008sg}. Given that our soliton bag has some features in
common with the magnetic bag description of a large number of
coincident $SU(2)$ magnetic monopoles, it could be interesting to
study both problems for higher rank gauge groups, as new features
certainly emerge for non-abelian monopoles beyond $SU(2)$.

In this paper we have restricted our investigations to the
effective five-dimensional Yang-Mills-Chern-Simons
version of the Sakai-Sugimoto model.
However, it would be a simple matter to employ our ansatz directly in
the full string theory version of the
Sakai-Sugimoto model, with the usual caveat that a
prescription must be employed to deal with the non-abelian Dirac-Born-Infeld
action. We expect the same kind of behaviour as in the five-dimensional
Yang-Mills-Chern-Simons version of the theory.

\section*{Appendix}\renewcommand{\thesection}{A}\setcounter{equation}{0}\quad
In this appendix we consider the Skyrme model and explain how the homogeneous approximation may be thought of as a smeared version of the Skyrme crystal. In particular, we show that the homogeneous approximation is an unattainable idealization that provides a lower bound on the energy of the true Skyrme crystal.

In terms of the hermitian currents $R_i=i(\partial_i U)U^{-1}$ the static energy
of the Skyrme model (in Skyrme units) is
\be
E=\int\bigg(
\frac{1}{2}\mbox{Tr}(R_iR_i)-\frac{1}{16}\mbox{Tr}([R_i,R_j][R_i,R_j])\bigg)\,d^3x
      \label{skyrmeenergy}
      \ee
      and the baryon number is
      \be
  B=\frac{i}{24\pi^2}\int \epsilon_{ijk}
  \mbox{Tr}(R_iR_jR_k)
    \,d^3x.    
 \label{skyrmebaryon}
 \ee
 The Faddeev-Bogomolny bound is $E\ge 12\pi^2 B$, but for non-zero $B$ this bound cannot be attained, as it requires that the Skyrme field $U({\bf x})$ is an isometry from $\mathbb{R}^3$ to $SU(2)$, which are two spaces that are not isometric.

 The solution of the Skyrme model that is closest to the bound is the triply periodic Skyrme crystal with $E/B=12\pi^2\times 1.04.$ This is a cubic lattice that contains four Skyrmions within a cube of side length $L=4.7$, and hence has a baryon number density ${\cal B}=4/L^3=0.04.$

 Substituting our idealized homogeneous approximation $R_i=-\beta\sigma_i/2$
 into (\ref{skyrmeenergy}) and (\ref{skyrmebaryon}) gives the energy per unit volume ${\cal E}$ and the baryon number density ${\cal B}$ to be
 \be
    {\cal E}=\frac{3}{4}\beta^2+\frac{3}{16}\beta^4, \qquad\qquad
    {\cal B}=\frac{\beta^3}{16\pi^2}.
    \ee
Using these expressions, we recover the energy bound by the simple manipulation
    \be
       {\cal E}=\frac{3}{4}\bigg(\beta-\frac{1}{2}\beta^2\bigg)^2+\frac{3}{4}\beta^3\ge \frac{3}{4}\beta^3=12\pi^2{\cal B}.
       \ee
Within the homogeneous approximation the bound is attained by $\beta=2$ with a corresponding baryon number density ${\cal B}=1/(2\pi^2)=0.05.$  These values provide a good estimate of both the energy and the baryon number density of the Skyrme crystal. The homogeneous approximation generates a lower bound for the true energy of the Skyrme crystal because there are no Skyrme fields that generate the idealized homogeneous currents required to attain the bound.      

\section*{Acknowledgements} 
This work is funded by the STFC grant ST/J000426/1 and an STFC PhD Studentship.
We thank Andreas Schmitt for useful discussions.

\end{document}